# Is the track-event theory of cell survival internally consistent?

## Sonwabile Arthur Ngcezu[1], Hans Rabus[2]

[1] Charlotte Maxeke Johannesburg Academic Hospital, Johannesburg, 2193, South Africa
[2] Physikalisch-Technische Bundesanstalt (PTB), 10587 Berlin, Germany

## Abstract

The "track event theory" (TET) has been developed in recent years as an alternative to the phenomenological linear-quadratic model for cell survival under exposure to ionizing radiation, particularly for heavy charged particles. The TET is based on a few simple model assumptions including the possibility to derive some of the model parameters from nanodosimetry. This work intends to carve out more clearly the basic assumptions behind the TET and to critically review the resulting mathematical model equations. It is demonstrated that the model assumptions of Poisson distribution and statistical independence of the frequency distributions of so-called one-track and two-track events follow from the Poisson distribution of the number of tracks affecting the considered target volume. It is also shown that the modified TET model equation used in the literature for consideration of repair is inconsistent with the model assumptions and requires an additional model parameter. Furthermore, the derivation of the model parameters from nanodosimetric properties of particle track structure is revealed to lead to a pure exponential dose dependence when the potentially large number of relevant nanometric target volumes inside a cell nucleus is accounted for.

*Keywords:*
Nanodosimetry, track structure, track event theory

## 1. Introduction

The so-called track event theory (TET) proposed by Besserer and Schneider is a model for predicting cell survival based on the induction of DNA double strand breaks (DSBs) by charged particle tracks (Besserer and Schneider 2015a, 2015b). The induction of pairs of DSBs within a considered target volume by a particle track is called "event". (In contrast to microdosimetric terminology where "track" and "event" and both referring to the statistically correlated occurrence of energy transfer points (Booz et al. 1983; Rossi and Zaider 1996; Lindborg and Waker 2017).) The respective target sizes are in the nanometer range and the parameters of the model have been related to nanodosimetry (Schneider et al. 2016, 2017, 2019).

In the first version of the TET (Besserer and Schneider 2015a), the basic biophysical model assumption was that a cell will be inactivated if at least two sub-lethal lesions in the form of DSBs are induced by direct radiation interaction with the DNA. If the two or more sub-lethal lesions are produced by a single track, this is called a one-track event (OTE). If a track produces exactly one sub-lethal lesion, this is called a two-track-event (TTE), as it requires at least two tracks interacting in the cell for its inactivation. The mathematical formulation of the model further involved the assumption that OTEs and TTEs are "statistically independent independent events in the terminology of nanodosimetry".

Apart from being a bit obscure, this statement seems paradox given that for a particular track an OTE and a TTE are disjoint alternatives and, hence, statistically dependent. It will be shown in this article that this contradiction is coming from the fact that the terms OTE and TTE were used in (Besserer and Schneider 2015a, 2015b; Schneider et al. 2019) in two different meanings. Namely, on the one hand, the effect of a particular track on a cell in the sense stated above, and, on the other hand, for the (multi-event) result of the irradiation on the cell. Their mathematical formulation is based on the first meaning of the terms.

In the second version of the TET (Besserer and Schneider 2015b), the model assumption was relaxed by including the possibility of DSB repair, such that cell inactivation occurs if there are unrepaired sub-lethal lesions. Repair was assumed to be of "second order", meaning that DNA repair changes the cell survival rate only for cells with exactly two sublethal lesions. It will be shown here that in this version of the theory, a fourth model parameter would be needed that has been omitted in the work of Besserer and Schneider (2015b).

This deficiency also applies to later papers, in which attempts were made to reduce the number of adjustable model parameters by deriving the ratio of the two model parameters (related to OTEs and TTEs) from chromatin geometry and nanodosimetric properties of ion tracks (Schneider et al. 2016, 2017).

A third version of the TET was proposed in (Schneider et al. 2019) where an explicit relation is given between the principal TET model parameters and nanodosimetric parameters of track structure. This relation was derived by considering OTEs and TTEs in microscopic "lethal interaction" volumes (LIVs) within which DSBs are induced in "basic interaction volumes" (BIVs). A BIV is assumed to be a sphere of 2 nm diameter that contains a DNA segment of 5 to 10 base pairs. The size of the (spherical) LIV was found to be dependent on radiation type and ranged from 5 nm diameter for carbon ions up to 35 nm for photons. It will be shown here that such small target volumes in conjunction with the original TET model assumptions leads to a pure exponential dose dependence of cell survival.

## 2. Original version of the track-event theory

### 2.1. Basic formulation of the track-event theory

The TET considers that a track can lead to the induction of a lethal lesion which is called a one-track event (OTE) and will lead to cell inactivation. A track can also induce a sub-lethal lesion (SLL) that can transform into a lethal lesion if a second such lesion is produced in the same target volume by a second track. This is called a two-track-event (TTE). It is understood, although not explicitly mentioned in (Besserer and Schneider 2015a), that a track can also lead to neither an OTE nor a TTE.

As the sublethal lesions are assumed to be DSBs (Besserer and Schneider 2015, 2015b; Schneider et al. 2016, 2017, 2019), a TTE is a track that produces exactly one DSB in the considered target volume, and a OTE is a track that produces two or more DSBs. In this work, this is taken into account by using a subscript "1" for quantities related to TTEs and a subscript "2+" for quantities related to OTEs. (It should be noted that in the first formulation of the TET, an OTE was identified with the production of exactly "two lethal DSBs on the same or different chromosomes" (Besserer and Schneider 2015a). But as it was further stated that a "cell survives when there is no OTE or at most one TTE", it is obvious that implicitly the case of more than two DSBs was included in the OTE definition.)

The model assumptions of the TET are thus as follows:

a) The mean numbers of OTEs, $x_{2+}$, and TTEs, $x_1$, in a target volume are proportional to absorbed dose $D$, i.e.

$$x_{2+} = pD \qquad x_1 = qD \qquad (1)$$

where $p$ and $q$ are model parameters.

b) The probability distributions $P_1(n_1)$ and $P_{2+}(n_{2+})$ of the numbers $n_1$ and $n_{2+}$ of TTEs and OTEs, respectively, in a considered target volume are following Poisson statistics with distribution parameters $x_1$ and $x_{2+}$, respectively.

c) The stochastic quantities $n_1$ and $n_{2+}$ are statistically independent. (Besserer and Schneider used the obscure phrase that OTEs and TTEs "are statistically independent events in the terminology of nanodosimetry". From their mathematical formulation, e.g. equation (3) in (Besserer and Schneider 2015a), it is evident, that they assume statistical independence in the conventional sense.) Therefore, the probability $P(n_1,n_2)$ for simultaneous occurrence of $n_{2+}$ OTEs and $n_1$ TTEs can be factorized:

$$P(n_1, n_{2+}) = P_1(n_1) \times P_{2+}(n_{2+}) \qquad (2)$$

d) A single DSB is always non-lethal.

e) Two or more DSBs always lead to cell death.

The model assumptions lead to the following basic model equation of the TET for the probability $S$ of cell survival (Besserer and Schneider 2015a):

$$S = (1 + qD)e^{-(p+q)D} \qquad (3)$$

A low dose approximation of this model equation was shown to be equivalent to the commonly used linear-quadratic (L-Q) model and to have a dose dependence that matches the experimentally observed exponential dose dependence at higher doses (Besserer and Schneider 2015a).

### 2.2. Critical observations on the TET model

In the previous section, we have tried to overcome some of the shortcomings of the original formulation of the TET model by Besserer and Schneider (2015a) in the form of ambiguous or inconsequentially used terminology. One example was already mentioned, namely their use of the term TTE for a track inducing a single sub-lethal lesion as well as for occurrence of two tracks inducing sub-lethal lesions that form a lethal lesion. Another example is that the illustration of the basic interactions considered in the model shown in Fig. 1 of (Besserer and Schneider 2015a) suggests that the possibility of more than one track affecting the target volume is considered, which reflects model assumption b). At the same time, cases like that there is only one track that induces only one sub-lethal lesion or that there are more than two sublethal interactions by a single track are seemingly not included. In our formulation of model assumption e) we have included the latter possibility, because otherwise the model assumptions would not cover all relevant cases and because in their mathematical formulation of the model, Besserer and Schneider implicitly subsumed the case of more than two sub-lethal lesions when talking about two sub-lethal lesions.

It should also be noted that the identification of a sub-lethal lesion with a DSB and a lethal lesion with two or more DSBs is questionable. The values for parameter $q$ shown in Table 1 of (Besserer and Schneider 2015a) suggest that around one DSB is induced per Gy of absorbed dose whereas evidence in radiobiological literature indicates that there are generally in the order of several tens per Gy (Ward 1990). However, DSB repair was not included in the first version of the TET and, furthermore, it may be necessary to consider only severe lesions in form of complex DSBs. Therefore, this aspect is not further investigated in this work that is focussed on the internal consistency of the TET.

With respect to internal consistency, it must be noted that model assumption c) is not immediately evident, given that TTE and OTE are distinct alternatives, i.e. mutually exclusive, for a particular track. As a track can either lead to an OTE or a TTE or to none of the two, the single event distributions of OTEs and TTEs are not statistically independent as the probability for simultaneous occurrence is zero. Both are Bernoulli distributed (only 0 or 1 as possible outcome) with the distribution parameters being the conditional probabilities for induction of an OTE or a TTE by a particular track. In the next section, we show that for multi-event distributions this assumed statistical independence applies, however.

### 2.3. Relation of TET model parameters and tracks

The model parameters $p$ and $q$ of the TET have a dimension of a reciprocal dose which masks their relation to the particle tracks in the radiation field. If we consider a particular irradiation and a cross section of the radiation beam, then the model parameters $p$ and $q$ are given by

$$p = \iint_A p_{c,2+}(\boldsymbol{r}) \frac{\Phi(\boldsymbol{r}|D)}{D} d^2\boldsymbol{r} \qquad (4)$$

$$q = \iint_A p_{c,1}(\mathbf{r}) \frac{\Phi(\mathbf{r}|D)}{D} d^2\mathbf{r} \quad (5)$$

In eqs. (4) and (5), $p_{c,1}(\mathbf{r})$ and $p_{c,2+}(\mathbf{r})$ are the conditional probabilities that a particle trajectory passing the considered plane at point $\mathbf{r}$ produces a TTE and an OTE, respectively, in the target volume; $\Phi(\mathbf{r}|D)$ is the dose-dependent area probability density (fluence) for a track passing at point $\mathbf{r}$. The integral extends over an area $A$ that is defined by the condition that tracks passing the beam cross section within this area have a non-zero probability for producing OTEs or TTEs in the considered target volume. If the occurrence of TTEs and OTEs is related to the formation of ionization clusters in particle tracks (Schneider et al. 2016, 2017, 2019), the diameter of area $A$ may be between several hundreds of nm up to more than a µm (Braunroth et al. 2020).

It should be noted that for the sake of avoiding the notation becoming too cumbersome, we ignore in eqs. (4) and (5) that $p_{c,1}(\mathbf{r})$ and $p_{c,2+}(\mathbf{r})$ also depend on the energy of the ionizing particle producing the track. We also do not consider explicitly that there is a dependence on the direction of motion. (In fact, the probabilities will mainly depend on the impact parameter of the track with respect to the target volume.)

The quantities $p_1$ and $p_{2+}$ given by

$$p_1 = \frac{1}{n_t} \iint_A p_{c,1}(\mathbf{r}) \Phi(\mathbf{r}|D) d^2\mathbf{r} \quad (6)$$

$$p_{2+} = \frac{1}{n_t} \iint_A p_{c,2+}(\mathbf{r}) \Phi(\mathbf{r}|D) d^2\mathbf{r} \quad (7)$$

are the (fluence-averaged) average probabilities that a track induces an OTE and a TTE, respectively, and may be assumed to be almost independent on $D$, whereas the dose dependence is included in average number of tracks $n_t$ passing area $A$.

$$n_t(D) = \iint_A \Phi(\mathbf{r}|D) d^2\mathbf{r} \quad (8)$$

In consequence, the two model parameters $p$ and $q$ can be written as follows:

$$p = \frac{n_t(D) \times p_{2+}}{D} \quad (9)$$

$$q = \frac{n_t(D) \times p_1}{D} \quad (10)$$

It should be noted that $n_t$ is generally not an integer number; it is the expectation of the probability distribution $P_t(n)$ of the number $n$ of tracks passing area $A$ that can produce OTEs and TTEs in the considered target volume. For a certain number $n$ of tracks passing $A$, the conditional probability $P_c(n_1, n_{2+}|n)$ for simultaneous induction of $n_1$ TTEs and $n_{2+}$ OTEs is given by a multinomial distribution.

$$P_c(n_1, n_{2+}|n) = \frac{n!}{n_0! n_1! n_{2+}!} p_0^{n_0} p_1^{n_1} p_{2+}^{n_{2+}} \quad (11)$$

where $p_0 = 1 - p_1 - p_{2+}$ and $n_0 = n - n_1 - n_{2+}$.

The probability $P(n_1, n_{2+})$ for $n_1$ TTEs and $n_{2+}$ OTEs to be produced is then given by:

$$P(n_1, n_{2+}) = \sum_n P_c(n_1, n_{2+}|n) P_t(n) \quad (12)$$

If $P_t(n)$ is a Poisson distribution (with $n_t$ as distribution parameter), $P(n_1, n_{2+})$ is obtained as

$$P(n_1, n_{2+}) = \frac{(n_t p_1)^{n_1} (n_t p_{2+})^{n_{2+}}}{n_1! n_{2+}!} e^{-n_t(p_1 + p_{2+})} \quad (13)$$

so that indeed the combined probability for $n_1$ TTEs and $n_{2+}$ OTEs can be written as the product of the marginal distributions that are thus statistically independent and are indeed Poisson distributions.

According to the TET biophysical model assumption, cell survival thus occurs if $n_1 \leq 1$ and $n_{2+} = 0$, i.e.

$$S = (1 + n_t p_1) e^{-n_t(p_1 + p_{2+})} \quad (14)$$

Using equations (9) and (10), eq. (14) can be rewritten to obtain the basic TET model formula given in eq. (3).

## 3. Extension of the TET for repair

In the second version of the TET, model assumption e) was modified and implicitly a further assumption f) was added (Besserer and Schneider 2015b):

e$_2$) If exactly two DNA lesions are induced either by one OTE or two TTEs, they are both repaired with a probability $R$.
f) More than two DNA lesions always lead to cell death.

In the respective model equation derived in (Besserer and Schneider 2015b) as eq. (7), the factor in front of the exponential in eq. (3) is replaced by a third-order polynomial in absorbed dose.

### 3.1. Critical observations on the TET model with repair

It is plausible that the repair capacity of a cell is limited so that for a large number of DSBs the average probability for an individual DSB to be repaired will decrease. However, it seems rather implausible that this should already be the case for three DSBs in a cell. In radiobiological assays, often a large number of DSB repair foci are observed (MacPhail et al. 2003; Ponomarev and Cucinotta 2006; Ponomarev et al. 2008; Martin et al. 2013). Hence, it might have been more appropriate to rather assume in the model a constant probability for repair of an individual DSB. Deriving a respective model equation becomes very intricate, however, as an analytical treatment of this case would require knowledge of the all probabilities $p_k$ for induction of $k$ DSBs by a single track.

A more critical issue, that has been ignored in the work of Besserer and Schneider (2015b) and later publications by Schneider et. al. (2016, 2017, 2019), is that model assumptions d) and e) imply the introduction of a fourth model parameter (in addition to the repair probability $R$).

The reason for this is that if only exactly two DSBs can be repaired, one has to make a distinction between tracks producing exactly two DSBs and those that produce three and more DSBs (see Fig. 1). The illustration in Fig. 1 of (Besserer and Schneider 2015b) suggests that the latter case was not considered.

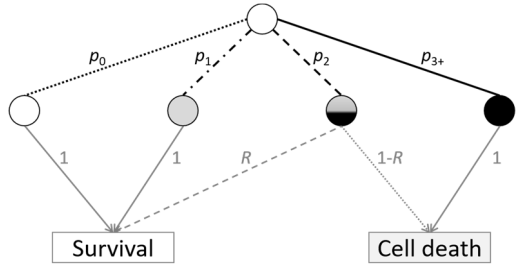

**Fig. 1** Illustration of a the fate of a cell interacting with a single track. The upper open circle symbolizes a cell without radiation damage. The interaction with the track may induce no DSB (dotted line), one DSB (dot-dashed line), two DSBs (dashed line) or more than two DSBs (solid line). In the first case, the cell remains without DSB (open circle) and survives. The second case leads to a cell with a DSB (grey circle) that is repaired with 100% probability (solid grey line). A cell with two DSBs (circle filled half with grey and black) has a limited probability to survive if both DSBs are repaired (dashed grey line) and else dies (dotted grey line). A cell with more than two DSBs dies at 100% probability.

If induction of two or more than two DSBs occurs with average probabilities $p_2$ and $p_{3+}$, respectively, the conditional probability $P_c(n_1, n_2, n_{3+}|n)$ for simultaneous occurrence of $n_1$, $n_2$ and $n_{3+}$ tracks inducing one, two and more than two DSBs in the considered target volume is given by:

$$P_c(n_1, n_2, n_{3+}|n) = \frac{n!\, p_0^{n_0} p_1^{n_1} p_2^{n_2} p_{3+}^{n_{3+}}}{n_0!\, n_1!\, n_2!\, n_{3+}!} \quad (15)$$

Weighting with the Poisson distribution of the number of tracks leads to the probability distribution $P'(n_1, n_2, n_{3+})$ as given in eq. (16).

$$P'(n_1, n_2, n_{3+}) = \frac{(n_t p_1)^{n_1} (n_t p_2)^{n_2} (n_t p_{3+})^{n_{3+}}}{n_1!\, n_2!\, n_{3+}!} e^{-n_t(p_1+p_2+p_{3+})} \quad (16)$$

According to model assumptions $e_2$) and f), a cell survives with probability $S$ given by

$$S = P'(0,0,0) + P'(1,0,0) + R[P'(0,1,0) + P'(2,0,0)] \quad (17)$$

Using eqs. (9) and (10), this transforms into

$$S = \left(1 + qD + R\left[p'D + \frac{(qD)^2}{2}\right]\right) e^{-(p+q)D} \quad (18)$$

where $p'$ is a fourth model parameter which is related to the probability that a track produces exactly two DSBs:

$$p' = \frac{n_t(D) \times p_2}{D} \quad (19)$$

Eq. (18) differs from the model equations given in (Besserer and Schneider 2015b; Schneider et al. 2017, 2019) by the absence of mixed terms (containing $p \times q$) and the absence of a term that is quadratic in the repair probability and cubic in dose. This is also true if instead of $e_2$) and f) one assumes that repair with probability $R$ occurs whenever there is more than one DSB. Then the probability for cell survival $S'$ would be given by eq. (20).

$$S' = (1 + qD)e^{-(p+q)D} + R[1 - (1 + qD)e^{-(p+q)D}] \quad (20)$$

The trivial reason is that the first term of the sum is the probability that at maximum one DSB is produced, so that the term in the square brackets is the probability for more than one DSB.

From eqs. (4) and (5) in (Besserer and Schneider 2015b), the mixed term (containing the product of $p$ and $q$) corresponds to the case of survival after two tracks interacted with the target volume; one track produces one DSB which is repaired with probability 1 and the other track two DSBs that both are repaired with probability $R$. The term quadratic in $R$ would correspond to three tracks of which one produces a pair of DSBs that are both repaired with probability $R$ while the other two tracks each produce a single DSB and the two DSBs coming from these two tracks are also repaired with a probability $R$.

From the point of view of DNA damage repair, there are two equivalent situations to the first case (mixed terms), namely one track that produces three DSBs or three tracks that each produce one. Similarly, the quadratic term involves four DSBs which would also be obtained by one track producing four DSBs or by one track producing three and a second track producing one or by two tracks producing two DSBs and also by the four tracks each producing one DSB. Hence, all these cases would have to be considered as well. However, this would require the respective probabilities to be used as further parameters of the model.

The reason why the approach of Besserer and Schneider (2015b) leads to the additional terms that are not appearing in eqs. (18) or (20) is that they implicitly assumed that the distributions of repaired DSBs produced by OTEs and TTEs would also be statistically independent if the numbers of OTEs and TTEs are statistically independent.

This is not the case, however, as the probability of repair should depend on the total number of DSBs produced and not how they are produced, as long as they are produced by tracks arriving with a time delay much smaller than the time needed for DSB repair. The latter is in the order of tens of minutes (Metzger and Iliakis 1991), so that for therapeutic beams, the DSBs produced by different tracks can be assumed to occur simultaneously.

The respective cell fate for the case of exactly two tracks is schematically illustrated in Fig. 2. The third row of circles shows the possible results of the interactions of the two tracks in the cell before repair. The possible results are no (open circle), one (grey circles), two (half grey and half black circles) and more than three (black circles). The solid grey lines indicate 100% probability, the dashed grey lines indicate repair with probability *R* and the dotted lines repair failure with probability (1-*R*).

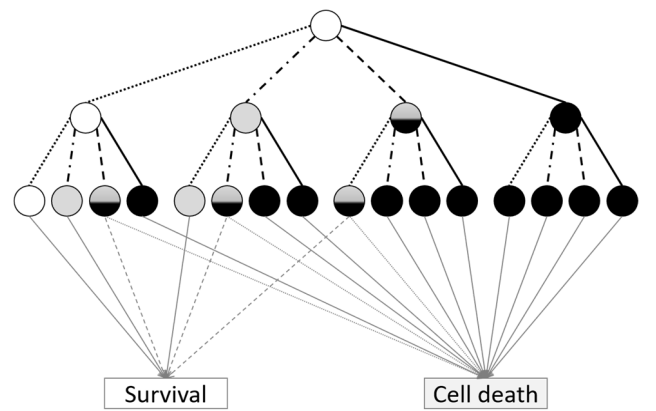

**Fig. 2** Illustration of the outcome when two tracks interact with a cell and DNA damage is repaired as per model assumptions e) and f). The meaning of the symbols and lines is the same as in Fig. 1.

In principle, Fig. 2 can also be seen as an illustration for more than two tracks interacting with the cell, if the second row of symbols is interpreted as the cell damage produced by all previous tracks where equivalent cases have been combined. As repair occurs at a much longer time scale than the production of the damage by the different tracks, there will not be quadratic terms in *R*. In summary, the considerations in this section mean that the model equations for the second version of the TET used in (Besserer and Schneider 2015b; Schneider et al. 2017, 2019) are incorrect or, at least, incompatible with the model assumptions.

## 4. TET and Nanodosimetry

In their further work, Schneider et. al. elaborated an approach to derive the ratio of model parameters *p* and *q* (Schneider et al. 2016, 2017) or the absolute parameter values (Schneider et al. 2019) from nanodosimetric parameters of track structure. To determine the absolute values of the parameters, they added the following model assumptions:

g) Existence of a certain nanometer sized "lethal interaction volume" (LIV) within which the induction of two or more (unrepaired) DSBs leads to cell death.
h) A DSB within the LIV is produced with a probability equal to the complementary cumulative probability $F_2$ for two or more ionizations within a "basic interaction volume" (BIV) that is traversed centrally by the primary particle trajectory.
i) The BIV is a sphere of about 2 nm diameter that encloses a short strand of DNA of 5 to 10 base pairs (Schneider et al. 2019).
j) The size of the LIV depends on radiation quality.

Based on these additional assumptions, the probability for an OTE (in the definition of the original model) and a TTE was then derived by binomial statistics and these values were finally used to derive an expression for RBE (Schneider et al. 2019).

### 4.1. Issues with the TET's link to nanodosimetry

This extension of the track-event model also has its deficiencies. The first is that eqs. (13) and (14) in (Schneider et al. 2019) give expressions for the model parameters *p* and *q* in form of probabilities, i.e. pure numbers in contrast to the dimension of a reciprocal dose (see eq. (7) in (Schneider et al. 2019)). Furthermore, they are exclusively derived from geometrical parameters and the single-track quantity $F_2$. Therefore, it is unclear how they could relate to a particular value of absorbed dose. (Unless one assumes them to relate to a particle fluence of 1 per cross section of the LIV, which would correspond to a dose in the order of kGy.) This deficiency could probably be healed using the approach presented in Section 2.3 and taking into account that a significant fraction of ionization clusters in a particle track are produced at radial distances of several tens to several hundreds of nm (Braunroth et al. 2020; Ngcezu et al. 2017; Rabus et al. 2020).

Another methodological problem is related to two aspects of model assumption h). The first aspect is that only central passage of the BIV is considered which ignores the fact mentioned in the previous sentence that tracks may produce ionization clusters in a target even if passing it at impact parameters in the order of hundreds of nm (Braunroth et al. 2020). In consequence, a better assumption would be that all BIVs in a LIV have the same probability for receiving an ionization cluster.

The second aspect is the assumed one-to-one correspondence between DSBs and the formation of ionization clusters of size two or more. While this has also been hypothesized in other work (Grosswendt 2005, 2006), comparisons with dedicated radiobiological experiments in work by Garty et al. showed the relation between ionization clusters and DSBs to require the use of a (one-parameter) combinatorial model (Garty et al. 2006; 2010), that was later demonstrated to imply the one-to-one correspondence between the probability for ionization clusters of two or more to apply only approximately and only for low-LET radiation (Nettelbeck and Rabus 2011; Rabus and Nettelbeck 2011). Conte et al. (Conte et al. 2017, 2018) and Selva et al. (Selva et al. 2019) demonstrated that a link between nanodosimetry and cell survival can be based on cumulative probabilities for ionization clusters if in addition to $F_2$ also the probability for ionization clusters of three or more, $F_3$, is included in the model.

However, a more severe problem lies in the fact that only one nanometric LIV is considered which gives up the advantage of the original TET model (where OTEs and TTEs were assumed to occur on the chromosome level) over aforementioned other attempts to link nanodosimetry and radiation biology (Grosswendt 2005, 2006; Garty et al. 2006, 2010; Conte et al. 2017, 2018; Selva et al. 2019; Schulte et al. 2008). As the LIV dimensions reported in (Schneider et al. 2019) are between 6 nm and 35 nm, such a volume covers only a small fraction of the volume of the cell nucleus on the order of $10^{-9}$ to $2\times10^{-7}$. Of course, one has to consider that DNA accounts only for a small fraction of the mass content in the nucleus and that, in addition, chromatin organization may play a role such that certain regions of the chromosome may be more prone to radiation damage (Schneider et al. 2016). However, even if there were only as few as 50 such sites per chromosome, the total number of LIVs in a cell nucleus would be on the order of $10^3$.

With more than one LIVs in the cell, model assumption f) implies that for each of them the condition must be fulfilled that no lethal damage occurs. If the single-LIV survival probability is given by eq. (3), then a cell will survive with a probability *S*

$$S = (1 + q_L D)^{n_L} e^{-n_L(p_L + q_L)D} \qquad (21)$$

where $n_L$ is the number of LIVs per cell nucleus and $p_L$ and $d_L$ are the respective model parameters for OTEs and TTEs occurring in a single LIV. For a large value of $n_L$, the first factor on the right-hand side of eq. (21) can be approximated by

$$(1 + q_L D)^{n_L} = \left(1 + \frac{n_L q_L D}{n_L}\right)^{n_L} \approx e^{n_L q_L D} \qquad (22)$$

so that finally the survival probability becomes

$$S = e^{-n_L p_L D} \qquad (23)$$

which is a pure exponential and describes the survival behavior for high-LET radiation (Goodhead et al. 1993).

## 5. Conclusions

The track event model was developed as an alternative model for the dose-dependence of cell survival that takes into account the radiation quality by including properties of particle track structure in form of nanodosimetric probabilities for ionization cluster formation. The original version of the TET produced a model equation (cf. eq. (3)) which offered the advantage of a functional shape that is equivalent to the L-Q model in the dose range where the latter describes the trend of experimental data well and is superior to it at higher doses where a pure exponential dose dependence is observed experimentally [1].

In this article, we have shown that some of the assumptions in the original model are dispensable: The Poisson statistics of the frequency distributions of OTEs and TTEs and their statistical independence can be derived from an assumed Poisson distribution of the number of tracks contributing to the numbers of OTEs and TTEs formed in the considered target volume.

As has been further shown in Sections 3 and 4, the later extensions of the TET are intrinsically inconsistent. With regard to the second version that considers repair, the problem was probably caused by the deceptive illustrations of the model shown as Fig. 1 of both papers of Besserer and Schneider (2015a, 2015b): These illustrations mix the definitions of OTEs and TTEs as the (multi-event) result of the irradiation on the cell and as the effect of a particular track on a cell. Therefore, important cases such as two tracks of which one produces two DSBs and the other one only one DSB are omitted that appear later implicitly in their model formula.

It seems that a better way to illustrate the model would have been decision trees (see Fig. 1 and Fig. 2) where the outcomes can be seen depending on the number of tracks interacting with the cell. Unfortunately, the fault in the mathematical model makes the comparison of the TET model with experimental data and the assessment of its performance with respect to the L-Q model more than questionable. Furthermore, the fact that the additional parameter $p'$ comes into play (see eq. (18)) counteracts the efforts to reduce the number of model parameters by estimating the ratio of model parameters $p$ and $q$ from nanodosimetric parameters of track structure.

It further appears that the consistent implementation of the model extension towards independent determination of some model parameters from nanodosimetric track properties leads to an exponential dose dependence. As this assessment relies on the statistical independence of lesion induction in different LIVs, recent experimental observations of correlations between ionization clusters in separate target volumes with a track (Pietrzak et al. 2018; Hilgers and Rabus 2020) give hope for a future substantially revised version of the TET that overcomes the deficiencies pointed out in this article.

## Acknowledgments

This work was supported by the German Federal Ministry for Economic Cooperation and Development (BMZ) in the frame of the Technical Cooperation Project "Upgrading of quality infrastructure in Africa". The National Metrology Institute of South Africa (NMISA) and the PTB Guest Researcher Programme are acknowledged for sponsoring guest researcher stays of S.A.N. at PTB.